%% file: DESY-02-217.tex
\documentclass[zpreprint,zbstpl]{./zeus_paper}
\usepackage[english]{babel}
\input ./zeus_def.tex
\input ./zeussubjetsdef.tex
\input  DESY-02-217-cit.tex

\begin{document}
\include{DESY-02-217-tit}

\include{authorslist_zeusc}

\include{DESY-02-217-txt}

\include{DESY-02-217-ref}
\include{DESY-02-217-tab}
\include{DESY-02-217-fig}

\end{document}

%% file: zeus_def.tex
\chardef\usc=95
\chardef\til=126
\catcode`\@=11 
\DeclareRobustCommand\xdotspace{\futurelet\@let@token\@xdotspace}
\def\@xdotspace{%
  \ifx\@let@token.\else
  \ifx\@let@token\bgroup.\else
  \ifx\@let@token\egroup.\else
  \ifx\@let@token\/.\else
  \ifx\@let@token\ .\else
  \ifx\@let@token~.\else
  \ifx\@let@token!.\else
  \ifx\@let@token,.\else
  \ifx\@let@token:.\else
  \ifx\@let@token;.\else
  \ifx\@let@token?.\else
  \ifx\@let@token/.\else
  \ifx\@let@token'.\else
  \ifx\@let@token).\else
  \ifx\@let@token-.\else
  \ifx\@let@token\@xobeysp.\else
  \ifx\@let@token\space.\else
  \ifx\@let@token\@sptoken.\else
   .\space
   \fi\fi\fi\fi\fi\fi\fi\fi\fi\fi\fi\fi\fi\fi\fi\fi\fi\fi}
\catcode`\@=12 

\newcommand{\stru}[2]{%
   \relax\ifmmode\hbox{\vrule height#1 depth#2 width0pt}%
   \else\vrule height#1 depth#2 width0pt\fi}

\newcommand{\Ronum}[1]{\uppercase\expandafter{\romannumeral#1}}
\newcommand{\ronum}[1]{\expandafter{\romannumeral#1}}
\DeclareRobustCommand{\LaTeXZ}{%
  \LaTeX\kern-.05em4\kern-.1em
  {\raisebox{-0.2ex}{$\scriptstyle\text{ZEUS}$}}\xspace}

\DeclareMathAlphabet{\mathbf}{OT1}{cmr}{bx}{sl}
\newcommand{\eVdist}{\kern-0.06667em}

\newcommand{\pb}{\,\text{pb}}

\newcommand{\slashfrac}[2]{%
  \raisebox{0.5ex}{\ensuremath #1}\kern-0.12em/\kern-0.08em
  \raisebox{-.8ex}{\ensuremath #2}}

\newcommand{\sqr}[3]{%
    {\vcenter{\hrule height.#3ex\hbox{\vrule width.#2ex height#1ex
     \kern#1ex\vrule width.#3ex}\hrule height.#2ex}}}

\catcode`\@=11 
\newcommand{\parenbar}{\mathpalette\p@renb@r}
\def\p@renb@r#1#2{\vbox{%
  \ifx#1\scriptscriptstyle \dimen@.7em\dimen@ii.2em\else
  \ifx#1\scriptstyle \dimen@.8em\dimen@ii.25em\else
  \dimen@1em\dimen@ii.4em\fi\fi \offinterlineskip
  \ialign{\hfill##\hfill\cr
    \vbox{\hrule width\dimen@ii}\cr
    \noalign{\vskip-.3ex}%
    \hbox to\dimen@{$\mathchar300\hfil\mathchar301$}\cr
    \noalign{\vskip-.3ex}%
    $#1#2$\cr}}}
\catcode`\@=12 

\newcommand{\IP}{{\rm I$\kern-0.01667em$P}\xspace}

\mathchardef\qsm=63
\mathchardef\pls=43
\mathchardef\mns=512
\mathchardef\plm=518
\mathchardef\eql=61
\mathchardef\smallleft=300
\mathchardef\smallright=301
\mathchardef\les=316
\mathchardef\gre=318
\mathchardef\leq=532
\mathchardef\grq=533

\catcode`\@=11 
\newcounter{pict@width}
\newcounter{pict@height}
\newlength{\pict@scale}
\newcommand{\psfigadd}[4]{%
\setcounter{pict@width}{1*\ratio{#2+\pict@scale/2}{\pict@scale}}
\setcounter{pict@height}{1*\ratio{#3+\pict@scale/2}{\pict@scale}}
\setlength{\unitlength}{\pict@scale}
\hbox to #2{\hspace{-\fill}\begin{picture}(\thepict@width,\thepict@height)
\put(0,0){\psfig{figure=#1,width=#2,height=#3,clip=}}
\SetScale{0.283466457}
\SetWidth{1.763889}
{#4}
\end{picture}}
}
\newcounter{pict@widthfst}
\newcounter{pict@widthscd}
\newcounter{pict@widthtot}
\newcommand{\psfigaddtwo}[7]{%
\setcounter{pict@widthfst}{1*\ratio{#2+\pict@scale/2}{\pict@scale}}
\setcounter{pict@widthscd}{1*\ratio{#2+#4+\pict@scale/2}{\pict@scale}}
\setcounter{pict@widthtot}{1*\ratio{#2+#4+#6+\pict@scale/2}{\pict@scale}}
\setcounter{pict@height}{1*\ratio{#3+\pict@scale/2}{\pict@scale}}
\setlength{\unitlength}{\pict@scale}
\hbox{\hspace{-\fill}\begin{picture}(\thepict@widthtot,\thepict@height)
\put(0,0){\psfig{figure=#1,width=#2,height=#3,clip=}}
\put(\thepict@widthscd,0){\psfig{figure=#5,width=#6,height=#3,clip=}}
\SetScale{0.283466457}
\SetWidth{1.763889}
{#7}
\end{picture}}
}
\newcommand{\psfigror}[4]{%
\setcounter{pict@width}{1*\ratio{#2+\pict@scale/2}{\pict@scale}}
\setcounter{pict@height}{1*\ratio{#3+\pict@scale/2}{\pict@scale}}
\setlength{\unitlength}{\pict@scale}
\hbox{\begin{picture}(\thepict@width,\thepict@height)
\put(0,\thepict@height){\psfig{figure=#1,width=#3,height=#2,clip=,angle=270}}
\SetScale{0.283466457}
\SetWidth{1.763889}
{#4}
\end{picture}}
}
\newcommand{\psfigrol}[4]{%
\setcounter{pict@width}{1*\ratio{#2+\pict@scale/2}{\pict@scale}}
\setcounter{pict@height}{1*\ratio{#3+\pict@scale/2}{\pict@scale}}
\setlength{\unitlength}{\pict@scale}
\hbox{\begin{picture}(\thepict@width,\thepict@height)
\put(0,0){\psfig{figure=#1,width=#3,height=#2,clip=,angle=90}}
\SetScale{0.283466457}
\SetWidth{1.763889}
{#4}
\end{picture}}
}
\catcode`\@=12 
\newlength\listtextwidth

\catcode`\@=11 
\newlength{\@tabfninsert}
\newlength{\@tabfnwidth}
\newcommand{\tabfootnote}[2]{%
  \setlength{\@tabfninsert}{0.8em}
  \setlength{\@tabfnwidth}{\textwidth}
  \addtolength{\@tabfnwidth}{-\@tabfninsert}
  \addtolength{\@tabfnwidth}{-0.4em}
  \noindent\makebox[\@tabfninsert][r]{\footnotesize$^{#1}$\hfil}\hfill%
  \parbox[t]{\@tabfnwidth}{\footnotesize #2\hfill}}
\catcode`\@=12 

%% file: zeussubjetsdef.tex
\def\pb1{pb$^{-1}$}

\def\kt{k_T}

\def\etjet{E_{T,\rm{jet}}}
\def\etajet{\eta_{\rm{jet}}}
\def\phijet{\phi_{\rm{jet}}}

\def\sq2{d\sigma/dQ^2}

\def\ptmiss{p_T\hspace{-4.2mm}\slash\hspace{1.5mm}}
\def\etaphi{\eta-\phi}

\def\eti{E_{T,i}}
\def\etai{\eta_i}
\def\phii{\phi_i}
\def\etj{E_{T,j}}
\def\etaj{\eta_j}
\def\phij{\phi_j}

%% file: DESY-02-217-cit.tex
\def\citeZEUS{{\cite{%
pl:b293:465,zeus:1993:bluebook%
}}\xspace}
\def\citeCTD{{\cite{%
nim:a279:290,*npps:b32:181,*nim:a338:254%
}}\xspace}
\def\citeCAL{{\cite{%
nim:a309:77,*nim:a309:101,*nim:a321:356,*nim:a336:23%
}}\xspace}

\def\calscale{pl:b531:9,*epj:c23:615,*hepex0206036}                        

\def\disent{np:b485:291}
\def\disaster{graudenz:1997,*hepph9710244}

\def\citeKTn{np:b406:187}
\def\citeKTesn{pr:d48:3160}

\def\citeSubjetsb{np:b383:419,*pl:b378:279}

\def\citesubmike{np:b421:545,*jhep:9909:009}

\def\sublep{zp:c63:363,*pl:b346:389,*pl:b374:304,*epj:c4:1,*epj:c17:1}
\def\subtevatron{pr:d65:052008}
\def\papelshapes{epj:c8:367}
\def\papelhighqcuad{epj:c11:427}

\def\papelas{pl:b507:70}

\def\sinistra{nim:a365:508,*nim:a391:360}

\def\dameth{proc:hera:1991:23b}

\def\snow{proc:snowmass:1990:134}

\def\lepto{cpc:101:108}

\def\heraclesb{cpc:69:155new,*spi:www:heracles}

\def\django{cpc:81:381,*spi:www:djangoh11}

\def\cdm{pl:b165:147,*pl:b175:453,*np:b306:746,*zfp:c43:625}
\def\ariadne{cpc:71:15,*zp:c65:285}

\def\lund{prep:97:31}
\def\jetset{cpc:39:347,*cpc:43:367}

\def\clustering{np:b238:492}
\def\herwig63{cpc:67:465,*jhep:0101:010,*hepph0107071}

\def\pdgnuevo{pr:d66:010001}

\def\cteqfour{pr:d55:1280}

\def\mrst99{epj:c4:463,*epj:c14:133}

%% file: DESY-02-217-tit.tex
\prepnum{{DESY--02--217}} 

\title{
Measurement of subjet multiplicities
in neutral current deep inelastic scattering
at HERA and determination of $\alpha_s$
}                                                       
                     
\author{ZEUS Collaboration}
\date{December, 2002}

\abstract{
    The subjet multiplicity has been measured in neutral current $e^+p$ interactions at
    $Q^2> 125$~GeV$^{\,2}$ with the ZEUS detector at HERA using an 
    integrated luminosity of $38.6$~pb$^{-1}$. Jets were identified in the
    laboratory frame using the longitudinally invariant $\kt$ cluster algorithm. 
    The number of jet-like substructures within jets, known as the subjet
    multiplicity,  
    is defined as the number of clusters resolved in a jet by reapplying the
    jet algorithm at a smaller resolution scale $y_{\rm cut}$. Measurements of the mean subjet
    multiplicity, $\bigl< n_{\rm sbj} \bigr>$, for jets with transverse energies $\etjet >15$~GeV
    are presented. Next-to-leading-order perturbative QCD calculations describe the
    measurements well. The value of $\alpha_s(M_Z)$, determined from 
    $\bigl< n_{\rm sbj}\bigr>$
    at $y_{\rm cut}=10^{-2}$ for jets with $25 < \etjet < 71$~GeV, is 
    $\alpha_s (M_Z) = 0.1187 \pm 0.0017 \; {\rm {(stat.)}}
    ^{+0.0024}_{-0.0009} \; {\rm {(syst.)}} ^{+0.0093}_{-0.0076} \; {\rm {(th.)}}$.
}

\makezeustitle

%% file: authorslist_zeusc.tex
\def\3{\ss}                                                                                        
\pagenumbering{Roman}                                                                              
                                                   %
\begin{center}                                                                                     
{                      \Large  The ZEUS Collaboration              }                               
\end{center}                                                                                       
  S.~Chekanov,                                                                                     
  D.~Krakauer,                                                                                     
  J.H.~Loizides$^{   1}$,                                                                          
  S.~Magill,                                                                                       
  B.~Musgrave,                                                                                     
  J.~Repond,                                                                                       
  R.~Yoshida\\                                                                                     
 {\it Argonne National Laboratory, Argonne, Illinois 60439-4815}~$^{n}$                            
\par \filbreak                                                                                     
  M.C.K.~Mattingly \\                                                                              
 {\it Andrews University, Berrien Springs, Michigan 49104-0380}                                    
\par \filbreak                                                                                     
  P.~Antonioli,                                                                                    
  G.~Bari,                                                                                         
  M.~Basile,                                                                                       
  L.~Bellagamba,                                                                                   
  D.~Boscherini,                                                                                   
  A.~Bruni,                                                                                        
  G.~Bruni,                                                                                        
  G.~Cara~Romeo,                                                                                   
  L.~Cifarelli,                                                                                    
  F.~Cindolo,                                                                                      
  A.~Contin,                                                                                       
  M.~Corradi,                                                                                      
  S.~De~Pasquale,                                                                                  
  P.~Giusti,                                                                                       
  G.~Iacobucci,                                                                                    
  A.~Margotti,                                                                                     
  R.~Nania,                                                                                        
  F.~Palmonari,                                                                                    
  A.~Pesci,                                                                                        
  G.~Sartorelli,                                                                                   
  A.~Zichichi  \\                                                                                  
  {\it University and INFN Bologna, Bologna, Italy}~$^{e}$
% ...
\newpage                                                                                           
$^{\    1}$ also affiliated with University College London \\                                      
% ...
\newpage   
                                                           %
                                                           %
\begin{tabular}[h]{rp{14cm}}                                                                       
$^{a}$ &  supported by the Natural Sciences and Engineering Research                               
          Council of Canada (NSERC) \\                                                             
$^{b}$ &  supported by the German Federal Ministry for Education and                               
          Research (BMBF), under contract numbers HZ1GUA 2, HZ1GUB 0, HZ1PDA 5, HZ1VFA 5\\         
$^{c}$ &  supported by the MINERVA Gesellschaft f\"ur Forschung GmbH, the                          
          Israel Science Foundation, the U.S.-Israel Binational Science                            
          Foundation and the Benozyio Center                                                       
          for High Energy Physics\\                                                                
$^{d}$ &  supported by the German-Israeli Foundation and the Israel Science                        
          Foundation\\                                                                             
$^{e}$ &  supported by the Italian National Institute for Nuclear Physics (INFN) \\                
$^{f}$ &  supported by the Japanese Ministry of Education, Science and                             
          Culture (the Monbusho) and its grants for Scientific Research\\                          
$^{g}$ &  supported by the Korean Ministry of Education and Korea Science                          
          and Engineering Foundation\\                                                             
$^{h}$ &  supported by the Netherlands Foundation for Research on Matter (FOM)\\                   
$^{i}$ &  supported by the Polish State Committee for Scientific Research,                         
          grant no. 620/E-77/SPUB-M/DESY/P-03/DZ 247/2000-2002\\                                   
$^{j}$ &  partially supported by the German Federal Ministry for Education                         
          and Research (BMBF)\\                                                                    
$^{k}$ &  supported by the Fund for Fundamental Research of Russian Ministry                       
          for Science and Edu\-cation and by the German Federal Ministry for                       
          Education and Research (BMBF)\\                                                          
$^{l}$ &  supported by the Spanish Ministry of Education and Science                               
          through funds provided by CICYT\\                                                        
$^{m}$ &  supported by the Particle Physics and Astronomy Research Council, UK\\                   
$^{n}$ &  supported by the US Department of Energy\\                                               
$^{o}$ &  supported by the US National Science Foundation\\                                        
$^{p}$ &  supported by the Polish State Committee for Scientific Research,                         
          grant no. 112/E-356/SPUB-M/DESY/P-03/DZ 301/2000-2002, 2 P03B 13922\\                    
$^{q}$ &  supported by the Polish State Committee for Scientific Research,                         
          grant no. 115/E-343/SPUB-M/DESY/P-03/DZ 121/2001-2002, 2 P03B 07022\\                    
\end{tabular}                                                                                      
                                                           %
                                                           %

%% file: DESY-02-217-txt.tex
\raggedbottom
\pagenumbering{arabic} 
\pagestyle{plain}
%
\section{Introduction}
\label{sec-int}

Jet production in $e^+p$ neutral current (NC) deep inelastic scattering (DIS)
provides a rich testing ground for perturbative QCD~(pQCD) and allows a precise
determination of the strong coupling constant, $\alpha_s$
\cite{pl:b363:201,\papelas,pl:b547:164,pl:b346:415,*epj:c5:625,*epj:c6:575,epj:c19:289}.
In the analysis described here, a new method is used to extract $\alpha_s$ in DIS,
which exploits the pQCD description of the internal structure of jets. The
investigation of such structure also gives information on the transition from a
parton produced in a hard subprocess to the experimentally observed jet of hadrons.
The method uses measurements of the mean subjet 
multiplicity for an inclusive sample of jets,
where the subjet multiplicity is defined as the number of clusters resolved in a jet
by reapplying the jet algorithm at a smaller resolution scale 
$y_{\rm cut}$~\cite{\citeSubjetsb,\citesubmike}.
At high transverse energy, $\etjet$, and for values of $y_{\rm cut}$ not too low,
fragmentation effects become small and the subjet multiplicity is calculable in pQCD.
Furthermore, the pQCD calculations depend only weakly on the knowledge of the parton
distribution functions (PDFs) of the proton, since the subjet multiplicity is determined
by QCD radiation processes in the final state. In zeroth order QCD a jet consists
of only one parton and the subjet multiplicity is trivially equal to unity. The first
non-trivial contribution to the subjet multiplicity is given by $\mathcal{O}(\alpha_s)$
processes in which, e.g., a quark radiates a gluon at a small angle. The deviation of 
the subjet multiplicity from unity is proportional to the rate of parton emission and
thus to $\alpha_s$. The next-to-leading-order (NLO) QCD corrections are
available, enabling $\alpha_s$
to be determined reliably. Measurements of subjet production
have been made in $e^+e^-$ interactions~\cite{\sublep}, $p\bar{p}$
collisions~\cite{\subtevatron} and NC DIS~\cite{np:b545:3} and have been used to
test the QCD predictions on coherence effects, differences between quarks and
gluons and splitting of jets.

This paper presents measurements of the mean subjet multiplicity in NC DIS
at $Q^2> 125$~GeV$^2$, where $Q^2$ is the virtuality of the exchanged boson, for an
inclusive sample of jets identified in the laboratory frame with the longitudinally
invariant $\kt$ cluster algorithm~\cite{\citeKTn,\citeKTesn}. The measurements are
compared to NLO QCD predictions~\cite{\disent} and are used to extract $\alpha_s(M_Z)$. 

\section{Experimental conditions}
\label{sec-exp}

The data sample was collected with the ZEUS detector at HERA and corresponds to an
integrated luminosity of $38.6 \pm 0.6$~\pb1. During 1996-97, HERA
operated with protons of energy $E_p=820$~GeV and positrons of energy $E_e=27.5$~GeV.
The ZEUS detector is described in detail elsewhere \citeZEUS. The main components used
in the present analysis are the central tracking detector (CTD)~\citeCTD, positioned
in a 1.43~T solenoidal magnetic field, and the uranium-scintillator sampling
calorimeter (CAL)~\citeCAL. The CTD was used to establish an interaction vertex with a
typical resolution along (transverse to) the beam direction of~$0.4$~($0.1$)~cm.

The CAL covers $99.7\%$ of the total solid angle. It is divided into three parts with a
corresponding division in the polar angle\footnote{The ZEUS coordinate system is a
right-handed Cartesian system, with the $Z$ axis pointing in the proton beam direction,
referred to as the ``forward direction'', and the $X$ axis pointing left towards the
centre of HERA. The coordinate origin is at the nominal interaction point. The
pseudorapidity is defined as $\eta=-\ln(\tan\frac{\theta}{2})$.}, $\theta$, as viewed
from the nominal interaction point: forward (FCAL, $2.6^{\circ}<\theta<36.7^{\circ}$),
barrel (BCAL, $36.7^{\circ}<\theta<129.1^{\circ}$), and rear (RCAL,
$129.1^{\circ}<\theta<176.2^{\circ}$). For normal incidence, the depth of the CAL is
seven interaction lengths in FCAL, five in BCAL and four in RCAL. Each of the calorimeter
parts is subdivided into towers which in turn are segmented longitudinally into one
electromagnetic (EMC) and one (RCAL) or two (FCAL, BCAL) hadronic (HAC) sections. The
FCAL and RCAL sections are further subdivided into cells with inner-face sizes of
$5 \times 20$~cm$^2$ ($10 \times 20$~cm$^2$ in the RCAL) for the EMC and 
$20 \times 20$~cm$^2$ for the HAC sections. The BCAL EMC cells have a projective
geometry as viewed from the nominal interaction point; each is $23.3$~cm long in the
azimuthal direction and has a width of $4.9$~cm along the beam direction at its inner
face, at a radius $123.2$~cm from the beam line. The BCAL HAC cells have a projective
geometry in the azimuthal direction only; the inner-face size of the inner (outer)
HAC section is $24.4 \times 27.1$~cm$^2$ ($24.4 \times 35.2$~cm$^2$). Each cell is
viewed by two photomultipliers. At $\theta = 90^{\circ}$, the size of an EMC (HAC)
cell in the pseudorapidity-azimuth ($\etaphi$) plane is approximately
$0.04 \times 11^{\circ}$ ($0.16 \times 11^{\circ}$). Under test-beam conditions, the
CAL energy resolution is $\sigma(E)/E = 18\%/\sqrt{E(\text{GeV})}$ for electrons and
$\sigma(E)/E = 35\%/\sqrt{E(\text{GeV})}$ for hadrons.
 
\section{Data selection and jet reconstruction}
\label{secsel}
 
A three-level trigger was used to select events online~\cite{zeus:1993:bluebook,trigger}.
The NC DIS events were selected offline using criteria similar to those
reported previously~\cite{pl:b547:164}. The main steps are outlined below.

The scattered-positron candidate was identified from the pattern of energy deposits
in the CAL \cite{\sinistra}. The energy ($E_{e}^{\prime}$) and polar angle
($\theta_{e}$) of the positron candidate were also determined from the CAL
measurements. The double angle method~\cite{\dameth}, which uses $\theta_{e}$ and an
angle~($\gamma$) that corresponds, in the quark-parton model, to the direction of the
scattered quark, was used to reconstruct $Q^2$ ($Q^2_{DA}$). The angle $\gamma$ was
reconstructed using the CAL measurements of the hadronic final state \cite{\dameth}.
The following requirements were imposed on the data sample:
\begin{itemize}
\item a positron candidate of energy $E_{e}^{\prime}>10$~GeV. This
     cut ensured a high and well understood positron-finding efficiency and suppressed
     background from photoproduction, in which the scattered positron escapes
     in the rear beampipe;
\item $y_e<0.95$, where $y_e=1-E_{e}^{\prime}(1-\cos{\theta_{e}})/(2 E_e)$.
     This condition removed events in which fake positron candidates from
     photoproduction background were found in the FCAL;
\item the energy not associated with the positron candidate within a cone of radius
     0.7 units in the $\etaphi$ plane around the positron direction was required
     to be less than $10\%$ of the positron energy. This condition removed
     photoproduction and DIS events in which part of a jet was incorrectly identified as
     the scattered positron;
\item for positrons in the polar-angle range $30^{\circ}<\theta_{e}<140^{\circ}$, the
     fraction of the positron energy within a cone of radius 0.3 units in the
     $\etaphi$ plane around the positron direction was required to be larger than
     0.9; for $\theta_{e}<30^{\circ}$, the cut was raised to 0.98. These
     requirements removed events in which a jet was incorrectly identified as the
     scattered positron;
\item the vertex position along the beam axis, determined from the CTD tracks,
     was required to be in the range $-38<Z<32$~cm, symmetrical around the mean
     interaction point for this running period;
\item $38<(E-p_Z)<65$~GeV, where $E$ is the total energy measured
     in the CAL, $E=\sum_iE_i$, and $p_Z$ is the $Z$ component of the
     vector ${\bf p}=\sum_i {E_i} \bf{r_i}$; in both cases the sum runs
     over all CAL cells, $E_i$ is the energy of the CAL cell $i$
     and ${\bf r_i}$ is a unit vector along the line joining the
     reconstructed vertex to the geometric centre of the cell $i$.
     This cut removed events with large initial-state radiation and further
     reduced the background from photoproduction;      
\item $\ptmiss/\sqrt{E_T}<2.5$~GeV$^{1/2}$, where $\ptmiss$ is the
     missing transverse momentum as measured with the CAL
     ($\ptmiss\equiv\sqrt{p_X^2+p_Y^2}$) and $E_T$ is the
     total transverse energy in the CAL. This cut removed
     cosmic rays and beam-related background;
\item events were rejected if a second positron candidate with energy above 10~GeV
     was found and the total energy in the CAL after subtracting that of the two
     positron candidates was below 4~GeV. This requirement removed elastic
     Compton-scattering events ($ep \rightarrow e\gamma p$);
\item $Q^2_{DA}>125$~GeV$^2$.
\end{itemize}

The longitudinally invariant $\kt$ cluster algorithm~\cite{\citeKTn} was used in the
inclusive mode~\cite{\citeKTesn} to reconstruct jets in the hadronic final state both
in data and in Monte Carlo (MC) simulated events (see Section~\ref{secmc}). In data,
the algorithm was applied in the laboratory frame to the energy deposits measured in
the CAL cells after excluding those associated with the scattered-positron candidate.
The jet search was performed in the $\etaphi$ plane. In the following discussion,
$\eti$ denotes the transverse energy, $\etai$ the pseudorapidity and $\phii$ the
azimuthal angle of object $i$. For each pair of objects (where the initial
objects are the energy deposits in the CAL cells), the quantity
\begin{equation}
 \label{dijeq}
 d_{ij} = [(\etai - \etaj)^2 + (\phii - \phij)^2 ] \cdot {\rm min}(\eti,\etj)^2
\end{equation}
was calculated. For each object, the quantity $d_i = (\eti)^2$
was also calculated. If, of all the values $\{d_{ij},d_i \}$, $d_{kl}$ was
the smallest, then objects $k$ and $l$ were combined into a single new
object. If, however, $d_k$ was the smallest, then object $k$ was considered
a jet and was removed from the sample. The procedure was repeated until all
objects were assigned to jets. The jet variables were defined according to the
Snowmass convention \cite{\snow}:
\begin{equation*}
 \etjet = \sum_i \eti \; ; \;
 \etajet = \frac{\sum_i \eti \etai}{\etjet} \; ; \;
 \phijet = \frac{\sum_i \eti \phii}{\etjet}.
\end{equation*}
This prescription was also used to determine the variables of the
intermediate objects.

Jet energies were corrected for all energy-loss effects, principally in inactive
material, typically about one radiation length, in front of the CAL. The corrected jet
variables were then used in applying additional cuts on the selected sample:
\begin{itemize}
\item events with at least one jet satisfying $\etjet>15$~GeV and $-1<\etajet<2$
     were selected;
\item events were removed from the sample if the distance of any of the jets to the
     positron candidate in the $\etaphi$ plane,
     \begin{equation*} 
      d = \sqrt{ (\etajet - \eta_e)^2 + (\phijet - \phi_e)^2},
     \end{equation*}
     was smaller than one unit. This requirement removed photoproduction background.
\end{itemize}
With the above criteria, 37~933~one-jet, 821~two-jet and 25~three-jet events 
were identified.
 
\subsection{Definition of the subjet multiplicity}

Subjets were resolved within a jet using all CAL cells associated with the
jet and repeating the application of the $\kt$ cluster algorithm described above,
until, for every pair of objects $i$ and $j$, the quantity $d_{ij}$ was greater than 
$d_{\rm cut}=y_{\rm cut}\cdot \big( \etjet \big)^2$~\cite{\citesubmike}. All remaining
objects were called subjets. The reconstruction of subjets within a jet was performed
using the uncorrected cell and jet energies, since systematic effects largely cancel
in the ratio $d_{ij}/ \big( \etjet \big)^2$ as seen in Eq.~(\ref{dijeq}).
The subjet structure depends upon the value chosen for the resolution parameter
$y_{\rm cut}$. The mean subjet multiplicity, $\bigl< n_{\rm sbj} \bigr>$, is defined
as the average number of subjets contained in a jet at a given value of $y_{\rm cut}$:
\begin{equation*}
 \bigl< n_{\rm sbj}(y_{\rm cut}) \bigr> = \frac{1}{N_{\rm jets}}\, \sum_{i=1}^{N_{\rm jets}}
                                    n^{i}_{\rm sbj}(y_{\rm cut}) \; ,
\end{equation*}
where $n^{i}_{\rm sbj}(y_{\rm cut})$ is the number of subjets in jet $i$ and
$N_{\rm jets}$ is the total number of jets in the sample. By definition,
$\bigl< n_{\rm sbj} \bigr> \geqslant 1$. The mean subjet multiplicity was measured for 
$y_{\rm cut}$ values in the range $5\cdot 10^{-4} - 0.1$. 

\section{Monte Carlo simulation}
\label{secmc}
 
Samples of events were generated to determine the response of the detector to jets of
hadrons and the correction factors necessary to obtain the hadron-level mean
subjet multiplicities. The generated events were passed through the
GEANT~3.13-based~\cite{tech:cern-dd-ee-84-1} ZEUS detector- and trigger-simulation
programs \cite{zeus:1993:bluebook}. They were reconstructed and analysed
by the same program chain as the data.
 
Neutral current DIS events were generated using the LEPTO~6.5 program \cite{\lepto}
interfaced to HERACLES~4.6.1 \cite{\heraclesb} via DJANGOH~1.1 \cite{\django}. The
HERACLES program includes photon and $Z$ exchanges and first-order electroweak
radiative corrections. The QCD cascade was modelled with the colour-dipole model
\cite{\cdm} by using the ARIADNE~4.08 program \cite{\ariadne} and including the
boson-gluon-fusion process. The colour-dipole model treats gluons emitted from
quark-antiquark (diquark) pairs as radiation from a colour dipole between two partons.
This results in partons that are not ordered in their transverse momenta. 
Samples of events were also generated using the model of LEPTO based on first-order
QCD matrix elements plus parton showers (MEPS). For the generation of the
samples with MEPS, the option for soft-colour interactions was switched
off~\cite{epj:c11:251}. In both cases, fragmentation into hadrons was performed using
the Lund \cite{\lund} string model as implemented in JETSET~7.4 \cite{\jetset}.
Events were also generated using the HERWIG 6.3~\cite{\herwig63} program, in which the
fragmentation into hadrons is simulated by a cluster model~\cite{\clustering}.
The CTEQ4D \cite{\cteqfour} proton PDFs were used for all simulations. 

The MC events were analysed with the same selection cuts and jet-search 
methods as were used for
the data. A good description of the measured distributions for the kinematic and jet
variables was given by both ARIADNE and LEPTO-MEPS. The simulations based on HERWIG
provided a poor description of the data at low values of $y_{\rm cut}$
($y_{\rm cut} \lesssim 5 \cdot 10^{-3}$) and, for this reason, it was not used to correct
the data. At relatively large values of $y_{\rm cut}$
($y_{\rm cut} \gtrsim 3 \cdot 10^{-2}$), HERWIG gave a good description of the data. The
identical jet algorithm was also applied to the hadrons (partons) to obtain predictions at
the hadron (parton) level. The MC programs were used to estimate QED radiative effects,
which were negligible for the measurements of $\bigl< n_{\rm sbj}\bigr>$.

\section{NLO QCD calculations}
\label{comnlo}

Experimental studies of QCD using jet production in NC DIS at HERA are often
performed in the Breit frame~\cite{bookfeynam:1972,*zfp:c2:237}. The analysis of the
subjet multiplicity presented here was instead performed in the laboratory frame,
since calculations of the mean subjet multiplicity for jets defined in the Breit frame
can, at present, only be performed to $\mathcal{O}(\alpha_s)$, precluding a reliable
determination of $\alpha_s$. However, calculations of the mean subjet multiplicity can
be performed up to $\mathcal{O}(\alpha_s^2)$ for jets defined in the laboratory frame. 
 
The perturbative QCD prediction for $\bigl< n_{\rm sbj} \bigr>$ was calculated as the
ratio of the cross section for subjet production to that for inclusive jet
production~($\sigma_{\rm jet}$):
\begin{equation}
\label{nsbjeq}
 \bigl< n_{\rm sbj} (y_{\rm cut})\bigr> = 1+\, \frac{1}{\sigma_{\rm jet}} \,
\sum_{j=2}^{\infty}(j-1) \cdot \sigma_{{\rm sbj},j}(y_{\rm cut}) \; ,
\end{equation}
where $\sigma_{{\rm sbj},j}(y_{\rm cut})$ is the cross section for producing jets
with $j$ subjets at a resolution scale of $y_{\rm cut}$. The NLO QCD predictions for
the mean subjet multiplicity were derived from Eq.~(\ref{nsbjeq}) by computing 
the subjet cross section to $\mathcal{O}(\alpha_s^2)$ and the inclusive jet cross
section to $\mathcal{O}(\alpha_s)$. As a result, the $\alpha_s$-dependence of the mean
subjet multiplicity up to $\mathcal{O}(\alpha_s^2)$ is given by
$\bigl< n_{\rm sbj} \bigr> = 1 + C_1 \ \alpha_s + C_2 \ \alpha_s^2$, where
$C_1$ and $C_2$ are quantities whose values depend on $y_{\rm cut}$ and the jet
and kinematic variables.

The measurements of the mean subjet multiplicity were performed in the
kinematic region defined by $Q^2> 125$~GeV$^2$ since, at lower values of $Q^2$, 
the sample of events with at least one jet with $\etjet > 15$~GeV is dominated by dijet
events. The calculation of the mean subjet multiplicity for dijet events can be
performed only up to $\mathcal{O}(\alpha_s)$, which would severely restrict the accuracy of
the predictions.

The measurements were compared with NLO QCD calculations using the program DISENT
\cite{\disent}. The calculations were performed in the $\overline{\rm MS}$ renormalisation
and factorisation schemes using a generalised version \cite{\disent} of the subtraction
method \cite{np:b178:421}. The number of flavours was set to five and the
renormalisation ($\mu_R$) and factorisation ($\mu_F$) scales were chosen to be
$\mu_R=\mu_F=Q$. The strong coupling constant, $\alpha_s$, was calculated at two loops
with $\Lambda^{(5)}_{\overline{\rm MS}}=202$~MeV, corresponding to $\alpha_s(M_{Z})=0.116$.
The calculations were performed using the CTEQ4M parameterisations of the proton PDFs.
The jet algorithm described in Section~\ref{secsel} was also applied to the partons in
the events generated by DISENT in order to compute the parton-level predictions for
the mean subjet multiplicity. The results obtained with DISENT were cross-checked by
using the program DISASTER++~\cite{\disaster}. The differences were smaller 
than $1\%$~\cite{thesisog}. Although DISENT does not include $Z$ exchange, its effect
in this analysis was negligible.

Since the measurements involve jets of hadrons, whereas the NLO QCD calculations
refer to partons, the predictions were corrected to the hadron level using
ARIADNE. The multiplicative correction factor, $C_{\rm had}$, was defined as the ratio
of $\bigl< n_{\rm sbj} \bigr>$ for jets of hadrons over that for jets of partons.
The value of $C_{\rm had}$ increases as $y_{\rm cut}$ decreases due to the increasing
importance of non-perturbative effects. The hadron-level prediction for
$\bigl< n_{\rm sbj} \bigr>$ approaches $\bigl< n^{\rm jet}_{\rm hadrons} \bigr>$ as
$y_{\rm cut}$ approaches $0$, where $\bigl< n^{\rm jet}_{\rm hadrons}\bigr>$ is the
mean multiplicity of hadrons in a jet. However, the maximum number of partons that
can be assigned to a jet in the NLO calculation is three, so the parton-level prediction
for $\bigl< n_{\rm sbj} \bigr>$ is restricted to $\bigl< n_{\rm sbj} \bigr> \leqslant 3$.
This fundamental problem was avoided by selecting high $\etjet$ and a relatively high
$y_{\rm cut}$ value, i.e. $\etjet > 25$~GeV and $y_{\rm cut} \geqslant 10^{-2}$.
In this region, the
hadronisation correction is small and the measured $\bigl< n_{\rm sbj} \bigr>$ is much
smaller than three, so that a reliable comparison of data and NLO QCD can be
made and $\alpha_s$ extracted. 

The procedure for applying hadronisation corrections to the NLO QCD calculations was
validated by verifying that the predicted dependence of the mean subjet multiplicity on
$y_{\rm cut}$ and $\etjet$ predicted by NLO QCD was well reproduced by both ARIADNE
and LEPTO-MEPS. The predictions based on HERWIG exhibited a different dependence both at
low values of $y_{\rm cut}$ and at high $\etjet$; for this reason, the HERWIG model
was not used in the evaluation of the uncertainty on the hadronisation correction.

The following sources were considered in the evaluation of the uncertainty affecting
the theoretical prediction of $\bigl< n_{\rm sbj} \bigr>$:
\begin{itemize}
 \item the uncertainty in the NLO QCD calculations due to terms beyond NLO, estimated by
      varying $\mu_R$ between $Q/2$ and $2Q$, was $\sim 3\%$ at $y_{\rm cut}=10^{-2}$;
 \item the uncertainty in the NLO QCD calculations due to that in the hadronisation
      correction was estimated as half of the difference between the values of
      $C_{\rm had}$ obtained with LEPTO-MEPS and with ARIADNE. It was smaller than $1.5\%$
      at $y_{\rm cut}=10^{-2}$ for $\etjet>25$~GeV;
 \item the uncertainty in the NLO QCD calculations due to the uncertainties in the
      proton PDFs was estimated by repeating the calculations using three additional
      sets of proton PDFs, MRST99, MRST99-g$\uparrow$ and
      MRST99-g$\downarrow$~\cite{\mrst99}. The differences were negligible;
 \item the NLO QCD calculations were carried out using
      $\mu_R=\etjet$ and $\mu_F=Q$. The differences were smaller than $0.3\%$
      at $y_{\rm cut}=10^{-2}$.
\end{itemize}

\section{Data corrections and systematic uncertainties}

The raw distribution of $n_{\rm sbj}$ in the data is compared to the prediction
of the ARIADNE simulation for several values of $y_{\rm cut}$ in Fig.~\ref{fig:nsbj}. 
The simulation provides a satisfactory description of the data, thus validating the use
of these MC samples to correct the measured mean subjet multiplicity to the hadron
level. Figure~\ref{fig:nsbj} also shows that the fraction of jets in the data with more
than three subjets at $y_{\rm cut}=10^{-2}$ is small; this fraction becomes negligible
for $\etjet > 25$~GeV, thus allowing a meaningful comparison with the NLO QCD
calculations. The 
mean subjet multiplicity corrected for detector
effects was determined bin-by-bin as 
$\bigl< n_{\rm sbj} \bigr>= K \bigl< n_{\rm sbj} \bigr>_{\rm CAL}$, 
where the correction factor was defined as 
$K=\bigl< n_{\rm sbj}\bigr>^{\rm MC}_{\rm had}/\bigl< n_{\rm sbj}\bigr>^{\rm MC}_{\rm CAL}$
and was evaluated separately for each value of $y_{\rm cut}$ in each region of
$\etjet$; the subscript ${\rm CAL}$ (${\rm had}$) indicates that
the mean subjet multiplicity was determined using the CAL cells (hadrons).
The deviation of the correction factor $K$ from unity was less
than $10\%$ for $y_{\rm cut} \geqslant 10^{-2}$ and decreased as $y_{\rm cut}$
increased.

The following sources of systematic uncertainty on the measurement of
$\bigl< n_{\rm sbj} \bigr>$ were considered~\cite{thesisog}:
\begin{itemize}
  \item the differences in the results obtained by using either ARIADNE or LEPTO-MEPS
       to correct the data for detector effects. This uncertainty was typically smaller
       than $1\%$;
  \item the scattered-positron candidate identification. The analysis was repeated by
       using an alternate technique~\cite{\papelhighqcuad} to select the
       scattered-positron candidate resulting in an uncertainty smaller than $0.5\%$;
  \item the $1\%$ uncertainty in the absolute energy scale of the jets~\cite{\calscale}
       resulted in an uncertainty smaller than $0.5\%$;
  \item the $1\%$ uncertainty in the absolute energy scale of the positron
       candidate~\cite{epj:c21:443} resulted in a negligible uncertainty;
  \item the uncertainty in the simulation of the trigger and in the cuts used to select
       the data also resulted in a negligible uncertainty.
\end{itemize}

\section{Measurement of the mean subjet multiplicity}
\label{secres}

The mean subjet multiplicity was measured for events with $Q^2 > 125$~GeV$^2$,
including every jet of hadrons in the event with $\etjet > 15$~GeV and
$-1 < \etajet < 2$, after correction for detector effects. It is shown as
a function of $y_{\rm cut}$ in Fig.~\ref{fig:ycutmc}a) and in Fig.~\ref{fig:ycutmc}b)
as a function of $\etjet$ at $y_{\rm cut}=10^{-2}$ and presented in Tables~\ref{tablansbjycut}
and \ref{tablansbjet}, respectively. The measured mean subjet multiplicity
decreases as $\etjet$ increases. This result is in agreement with that of a previous
publication~\cite{\papelshapes}, in which the internal structure of jets in NC DIS was
studied using the jet shape and it was observed that the jets become narrower as
$\etjet$ increases. This tendency is also consistent with the transverse-energy
dependence of the mean subjet multiplicity for jets identified in the Breit
frame~\cite{np:b545:3}.
 
The measurements in Fig.~\ref{fig:ycutmc} are compared with the predictions of the
ARIADNE and LEPTO-MEPS. The LEPTO-MEPS predictions overestimate the observed mean subjet
multiplicity; ARIADNE overestimates the data at low $\etjet$ and approaches the data 
at high $\etjet$.

Calculations of $\bigl< n_{\rm sbj} \bigr>$ in NLO QCD, corrected for hadronisation
effects, using the sets of proton PDFs of the CTEQ4 ``A-series'' are
compared to the data in Figs.~\ref{fig:ycutnlo} and~\ref{fig:etnlo}. The hadronisation
correction is small in the unshaded regions: as a function of $y_{\rm cut}$ and for
jets with $\etjet > 15$~GeV, $C_{\rm had}$ differs from unity by less than $25\%$ for
$y_{\rm cut} \geqslant 10^{-2}$ (see Fig.~\ref{fig:ycutnlo}); as a function of $\etjet$ at
$y_{\rm cut}=10^{-2}$, $C_{\rm had}$ differs from unity by less than $17\%$ for
$\etjet > 25$~GeV (see Fig.~\ref{fig:etnlo}). The measured $\bigl< n_{\rm sbj} \bigr>$
as a function of $y_{\rm cut}$ is well described by the NLO QCD predictions. For very
small $y_{\rm cut}$ values, the agreement is also good. In that region, fixed-order QCD
calculations are affected by large uncertainties and a resummation of terms
enhanced by $\ln{y_{\rm cut}}$~\cite{\citesubmike} would be required for a precise
comparison with the data. At relatively large values of $y_{\rm cut}$, an
NLO fixed-order calculation is expected~\cite{\citesubmike} to be a good
approximation to such a resummed calculation.

The sensitivity of the measurements to the value of $\alpha_s (M_Z)$ is illustrated in
Fig.~\ref{fig:etnlo} by the comparison of the measured $\bigl< n_{\rm sbj} \bigr>$ at 
$y_{\rm cut}=10^{-2}$ as a function of $\etjet$ with NLO QCD calculations for
different values of $\alpha_s(M_Z)$. The overall description of the data by the NLO QCD
calculations is good, so that the measurements can be used to make a determination of
$\alpha_s$.

\section{Determination of $\alpha_s$}
\label{secalphas}
 
The measurements of $\bigl< n_{\rm sbj} \bigr>$ for $25 < \etjet < 71$~GeV at
$y_{\rm cut}=10^{-2}$ were used to determine $\alpha_s(M_Z)$. The $y_{\rm cut}$ value
and the lower $\etjet$ limit were justified in Section~\ref{comnlo}; the value of
$C_{\rm had}$ differs from unity by less than $17\%$ and approaches unity as $\etjet$
increases. The mean value of $Q^2$ was $\bigl< Q^2 \bigr> = 1580$~GeV$^2$. The
following procedure was used:
\begin{itemize}
  \item NLO QCD calculations of $\bigl< n_{\rm sbj} \bigr>$ were performed for the
       five sets of the CTEQ4 ``A-series''. The value of $\alpha_s(M_Z)$ used in each
       partonic cross-section calculation was that associated with the corresponding
       set of PDFs;
  \item for each bin, $i$, in $\etjet$, the NLO QCD calculations, corrected for
       hadronisation effects, were used to parameterise the $\alpha_s(M_Z)$ dependence
       of $\bigl< n_{\rm sbj} \bigr>$ according to
    \begin{equation}
     \label{e:fit_as}
     \left[ \bigl< n_{\rm sbj} \bigr> (\alpha_s(M_Z))
       \right]_i = 1 + C_1^i \,\alpha_s(M_Z)+ C_2^i \,\alpha_s^2(M_Z) \;.
    \end{equation}
       The coefficients $C_1^i$ and $C_2^i$ were determined by performing a
       $\chi^2$-fit of this form to the NLO QCD predictions. The NLO QCD
       calculations were performed with an accuracy such that the statistical 
       uncertainties of these coefficients were negligible compared to any
       other uncertainty. This simple parameterisation gives a good description of the
       $\alpha_s(M_Z)$ dependence of $\bigl< n_{\rm sbj} \bigr>$ over the entire range
       spanned by the CTEQ4 ``A-series'';
  \item the value of $\alpha_s(M_Z)$ was then determined by a $\chi^2$-fit
       of Eq.~(\ref{e:fit_as}) to the measurements of $\bigl< n_{\rm sbj} \bigr>$.
       The resulting fit described the data well, giving $\chi^2=2.7$ for four 
       degrees of freedom.
\end{itemize}
 
This procedure correctly handles the complete $\alpha_s$-dependence of the NLO
calculations (the explicit dependence coming from the partonic cross sections and the
implicit one coming from the PDFs) in the fit, while preserving the correlation
between $\alpha_s$ and the PDFs.
 
The uncertainty on the extracted value of $\alpha_s (M_Z)$ due to the experimental
systematic uncertainties was evaluated by repeating the analysis above for each
systematic check. The largest contribution to the experimental uncertainty was that 
due to the simulation of the hadronic final state. A total systematic uncertainty
on $\alpha_s (M_Z)$ of $\Delta \alpha_s (M_Z) = {}^{+0.0024}_{-0.0009}$ was obtained
by adding in quadrature the individual contributions.
 
The theoretical uncertainties on $\alpha_s(M_Z)$ arising from terms beyond NLO and uncertainties in the hadronisation
correction, evaluated as described in
Section~\ref{comnlo},  were found to be $\Delta \alpha_s (M_Z) = ^{+0.0089}_{-0.0071}$ and
$\Delta \alpha_s (M_Z)= \pm 0.0028$, respectively. The total theoretical uncertainty was
obtained by adding these uncertainties in quadrature. The results are presented in
Table~\ref{tablaas}. In addition, as a cross check, the
measurement was repeated using three of the MRST99 sets of proton PDFs: central,
$\alpha_s \uparrow\uparrow$ and $\alpha_s \downarrow\downarrow$. The result agreed
with that obtained by using CTEQ4 to better than $0.3\%$. The value 
of $\alpha_s$ is
in agreement with the central result for variations in the choice of $y_{\rm cut}$ 
in the range $5\cdot 10^{-3}$ to $3\cdot 10^{-2}$.

The value of $\alpha_s(M_Z)$ as determined from the measurements of
$\bigl< n_{\rm sbj} \bigr>$ for $25 < \etjet < 71$~GeV at $y_{\rm cut}=10^{-2}$ is
\begin{equation}
\alpha_s (M_Z) = 0.1187 \pm 0.0017 \; {\rm {(stat.)}}
^{+0.0024}_{-0.0009} \; {\rm {(syst.)}} ^{+0.0093}_{-0.0076} \; {\rm {(th.)}}\;. \nonumber
\end{equation}
This result is consistent with recent determinations by the
H1~\cite{epj:c19:289,epj:c21:33} and ZEUS~\cite{\papelas,pl:b547:164,mandyfit}
Collaborations and with the PDG value, $\alpha_s(M_Z)=0.1172 \pm 0.0020$~\cite{\pdgnuevo}.
This determination of $\alpha_s$ has experimental uncertainties as small as those
based on the measurements of jet cross sections in DIS. However, the theoretical
uncertainty is larger and dominated by terms beyond NLO. Further theoretical work on
higher-order contributions would allow an improved measurement.

\section{Summary}
\label{secsumm}     

 Measurements of the mean subjet multiplicity for jets produced in neutral current
deep inelastic $e^+p$ scattering at a centre-of-mass energy of 300~GeV have been
made using every jet of hadrons with $\etjet > 15$~GeV and $-1 < \etajet < 2$
identified with the longitudinally invariant $\kt$ cluster algorithm in the laboratory
frame. The average number of subjets within a jet decreases as $\etjet$ increases.

Next-to-leading-order QCD calculations reproduce the measured values well,
demonstrating a good description of the internal structure of jets by QCD radiation.
The mean subjet multiplicity of an inclusive sample of jets produced in NC DIS has
the advantage of being mostly sensitive to final-state parton-radiation processes and
of allowing an extraction of $\alpha_s$ with very little dependence on the proton
parton distribution functions.

A QCD fit of the measurements of the mean subjet multiplicity for $25 < \etjet < 71$~GeV
at $y_{\rm cut}=10^{-2}$ yields
\begin{displaymath}
 \alpha_s (M_Z) = 0.1187 \pm 0.0017 \; {\rm {(stat.)}}
 ^{+0.0024}_{-0.0009} \; {\rm {(syst.)}} ^{+0.0093}_{-0.0076} \; {\rm {(th.)}}\;.
\end{displaymath}

\vspace{0.5cm}
\noindent {\Large\bf Acknowledgments}
\vspace{0.3cm}
 
We thank the DESY Directorate for their strong support and encouragement.
The remarkable achievements of the HERA machine group were essential for
the successful completion of this work and are greatly appreciated. We
are grateful for the support of the DESY computing and network services.
The design, construction and installation of the ZEUS detector have been
made possible owing to the ingenuity and effort of many people from DESY
and home institutes who are not listed as authors. We would like to thank
M. Seymour for valuable discussions.
 
\vfill\eject

%% file: DESY-02-217-ref.tex
\pagestyle{plain}
{\raggedright
\providecommand{\etal}{et al.\xspace}
\providecommand{\coll}{Collaboration}
\catcode`\@=11
\def\@bibitem#1{%
\ifmc@bstsupport
  \mc@iftail{#1}%
    {;\newline\ignorespaces}%
    {\ifmc@first\else.\fi\orig@bibitem{#1}}
  \mc@firstfalse
\else
  \mc@iftail{#1}%
    {\ignorespaces}%
    {\orig@bibitem{#1}}%
\fi}%
\catcode`\@=12

}
\vfill\eject

%% file: DESY-02-217-tab.tex

\begin{table}[p]
\begin{center}
\begin{tabular}{|c|ccc||c|}
\hline
\raisebox{0.25cm}[1.cm]{\parbox{2cm}{\centerline{$y_{cut}$ value}}}
          & \raisebox{0.25cm}[1.cm]{\parbox{2cm}{\centerline{$\big< n_{sbj} \big>$}}}
                           & \raisebox{0.25cm}[1.cm]{$\Delta_{stat}$} &
                             \raisebox{0.25cm}[1.cm]{$\Delta_{syst}$} &
           \raisebox{0.2cm}[0.8cm]{\parbox{2.8cm}{\centerline{PAR to HAD} \vspace{-0.2cm} \centerline{correction}}}\\[.05cm]
\hline
   &&&&\\[-.4cm]
   $0.0005$ &
              $4.3013$  & $\pm 0.0068$ &  $^{+0.0363}_{-0.0031}$ &  
                           $1.981 \pm 0.090$
   \\[.2cm]
   $0.001$ &
              $3.4975$  & $\pm 0.0057$ &  $^{+0.0306}_{-0.0037}$ &  
                           $1.744 \pm 0.073$
   \\[.2cm]
   $0.003$ &
              $2.4457$  & $\pm 0.0043$ &  $^{+0.0260}_{-0.0039}$ &  
                           $1.487 \pm 0.056$
   \\[.2cm]
   $0.005$ &
              $2.0630$  & $\pm 0.0038$ &  $^{+0.0175}_{-0.0040}$ &  
                           $1.387 \pm 0.047$
   \\[.2cm]
   $0.01$ &
              $1.6347$  & $\pm 0.0033$ &  $^{+0.0111}_{-0.0034}$ &  
                           $1.246 \pm 0.029$
   \\[.2cm]
   $0.03$ &
              $1.2015$  & $\pm 0.0024$ &  $^{+0.0024}_{-0.0061}$ &  
                           $1.0651 \pm 0.0057$
   \\[.2cm]
   $0.05$ &
              $1.1019$  & $\pm 0.0018$ &  $^{+0.0014}_{-0.0040}$ &  
                           $1.0224 \pm 0.0018$
   \\[.2cm]
   $0.1$ &
              $1.0326$  & $\pm 0.0010$ &  $^{+0.0003}_{-0.0009}$ &  
                           $1.0015 \pm 0.0001$
   \\[.2cm]
\hline
\end{tabular}
\caption{
Measurement of the mean subjet multiplicity as a function of $y_{\rm cut}$.
The statistical
and systematic uncertainties are shown separately. The multiplicative correction
applied to correct for hadronisation effects is shown
in the last column.
}
  \label{tablansbjycut}
\end{center}
\end{table}
\begin{table}[p]
\begin{center}
\begin{tabular}{|c|ccc||c|}
\hline
\raisebox{0.25cm}[1.cm]{\parbox{2.6cm}{\centerline{$E_{T,{\rm jet}}$ interval}\centerline{(GeV)}}}
          & \raisebox{0.25cm}[1.cm]{\parbox{2cm}{\centerline{$\big< n_{sbj} \big>$}}} 
                           & \raisebox{0.25cm}[1.cm]{$\Delta_{stat}$} &
                             \raisebox{0.25cm}[1.cm]{$\Delta_{syst}$} &
           \raisebox{0.2cm}[0.8cm]{\parbox{2.8cm}{\centerline{PAR to HAD} \vspace{-0.2cm} \centerline{correction}}}\\[.05cm]
\hline
   &&&&\\[-.4cm]
   $15\;-\;17$ &
              $1.7792$  & $\pm 0.0060$ &  $^{+0.0122}_{-0.0052}$ &  
                           $1.332 \pm 0.043$
   \\[.2cm]
   $17\;-\;21$ &
              $1.6712$  & $\pm 0.0056$ &  $^{+0.0085}_{-0.0034}$ &  
                           $1.272 \pm 0.032$
   \\[.2cm]
   $21\;-\;25$ &
              $1.5654$  & $\pm 0.0079$ &  $^{+0.0186}_{-0.0050}$ &  
                           $1.198 \pm 0.020$
   \\[.2cm]
   $25\;-\;29$ &
              $1.481$  & $\pm 0.011$ &  $^{+0.022}_{-0.004}$ &  
                           $1.145 \pm 0.014$
   \\[.2cm]
   $29\;-\;35$ &
              $1.395$  & $\pm 0.012$ &  $^{+0.005}_{-0.003}$ &  
                           $1.0945 \pm 0.0055$
   \\[.2cm]
   $35\;-\;41$ &
              $1.333$  & $\pm 0.017$ &  $^{+0.007}_{-0.009}$ &  
                           $1.0536 \pm 0.0028$
   \\[.2cm]
   $41\;-\;55$ &
              $1.286$  & $\pm 0.019$ &  $^{+0.005}_{-0.011}$ &  
                           $1.0248 \pm 0.0022$
   \\[.2cm]
   $55\;-\;71$ &
              $1.293$  & $\pm 0.038$ &  $^{+0.004}_{-0.032}$ &  
                           $1.0049 \pm 0.0019$
   \\[.2cm]
\hline
\end{tabular}
\caption{
Measurement of the mean subjet multiplicity at $y_{\rm cut}=10^{-2}$
as a function of $\etjet$.
The statistical
and systematic uncertainties are shown separately. The multiplicative correction
applied to correct for hadronisation effects is shown
in the last column.
}
  \label{tablansbjet}
\end{center}
\end{table}
\begin{table}[p]
\begin{center}
\begin{tabular}{|c|cccc|}
\hline
\raisebox{0.25cm}[1.0cm]{\parbox{2.3cm}{\centerline{$E_{T,{\rm jet}}$ region} \centerline{(GeV)}}}            & \raisebox{0.15cm}[1.cm]{\parbox{2cm}{\centerline{$\alpha_s (M_Z)$}}}
                           & \raisebox{0.15cm}[1.cm]{$\Delta_{stat}$} &
                             \raisebox{0.15cm}[1.cm]{$\Delta_{syst}$} &
                             \raisebox{0.15cm}[1.cm]{$\Delta_{th}$} \\[0.05cm]
\hline
&&&&\\[-.4cm]
   $25\;-\;29$ &
              $0.1180$  & $\pm 0.0024$ &  $^{+0.0049}_{-0.0010}$ &  $^{+0.0098}_{-0.0080}$
   \\[.2cm]
   $29\;-\;35$ &
              $0.1181$  &  $^{+0.0031}_{-0.0032}$ & $^{+0.0014}_{-0.0007}$ & $^{+0.0088}_{-0.0071}$
   \\[.2cm]
   $35\;-\;41$ &
              $0.1191$ &  $^{+0.0050}_{-0.0051}$ & $^{+0.0021}_{-0.0025}$ & $^{+0.0088}_{-0.0072}$
   \\[.2cm]
   $41\;-\;55$ &
              $0.1207$ &  $^{+0.0060}_{-0.0062}$ & $^{+0.0017}_{-0.0036}$ & $^{+0.0087}_{-0.0072}$ 
   \\[.2cm]
   $55\;-\;71$ &
              $0.140$ &  $^{+0.013}_{-0.014}$ & $^{+0.001}_{-0.012}$ & $^{+0.012}_{-0.010}$
   \\[.2cm]   
\hline
&&&&\\[-.4cm]
   $25 - 71$ &
              $0.1187$ &  $\pm 0.0017$ & $^{+0.0024}_{-0.0009}$ & $^{+0.0093}_{-0.0076}$ 
   \\[.2cm]
\hline
\end{tabular}
\caption{
The $\alpha_s (M_Z)$ values as determined from the QCD fit to the
measured $\big< n_{sbj} \big>$ at $y_{cut}=10^{-2}$ as a function of $E_{T,{\rm jet}}$, as well as
that obtained by combining all regions. The statistical, systematic and
theoretical uncertainties are shown separately.
}
  \label{tablaas}
\end{center}
\end{table}
%

%% file: DESY-02-217-fig.tex

\begin{figure}[p]
\setlength{\unitlength}{1.0cm}
\begin{picture} (19.0,18.5)
\put (1.5,0.0){\epsfig{figure=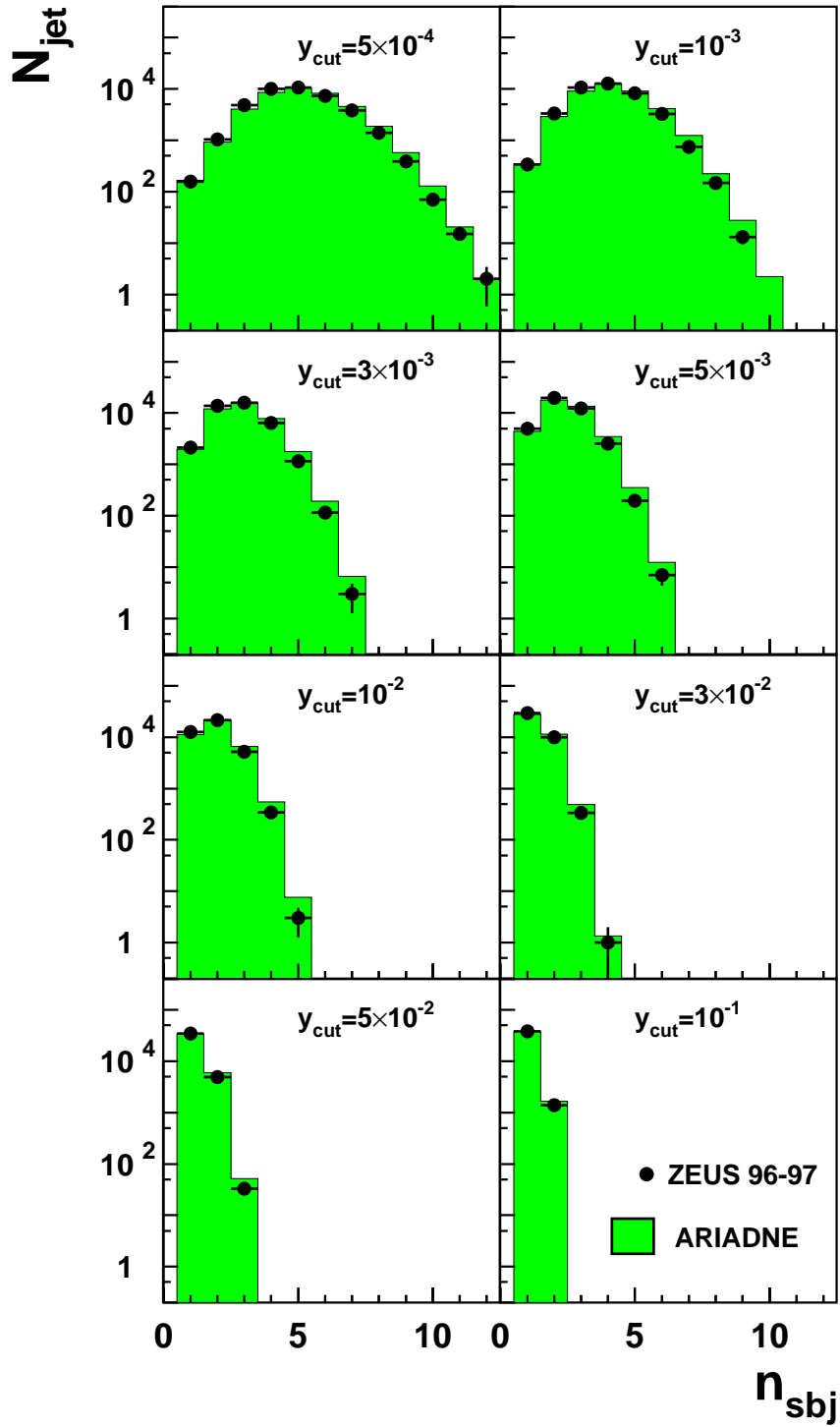,width=13.5cm}}
\end{picture}
\vspace{-0.5cm}
\caption{\label{fig:nsbj}
Distribution of the number of subjets within a jet at different 
values of $y_{\rm cut}$ 
for the inclusive sample of jets with $\etjet>15$~GeV and $-1<\etajet<2$ in NC
DIS at \mbox{$Q^2>125$~GeV$^{\,2}$}~(dots). The error bars show
the statistical uncertainty. For comparison, the predictions of
the ARIADNE simulation, area normalised to the data, are also shown as the histograms.
}
\end{figure}
\begin{figure}[p]
\setlength{\unitlength}{1.0cm}
\begin{picture} (19.0,18.50)
\put (1.20,7.4){\epsfig{figure=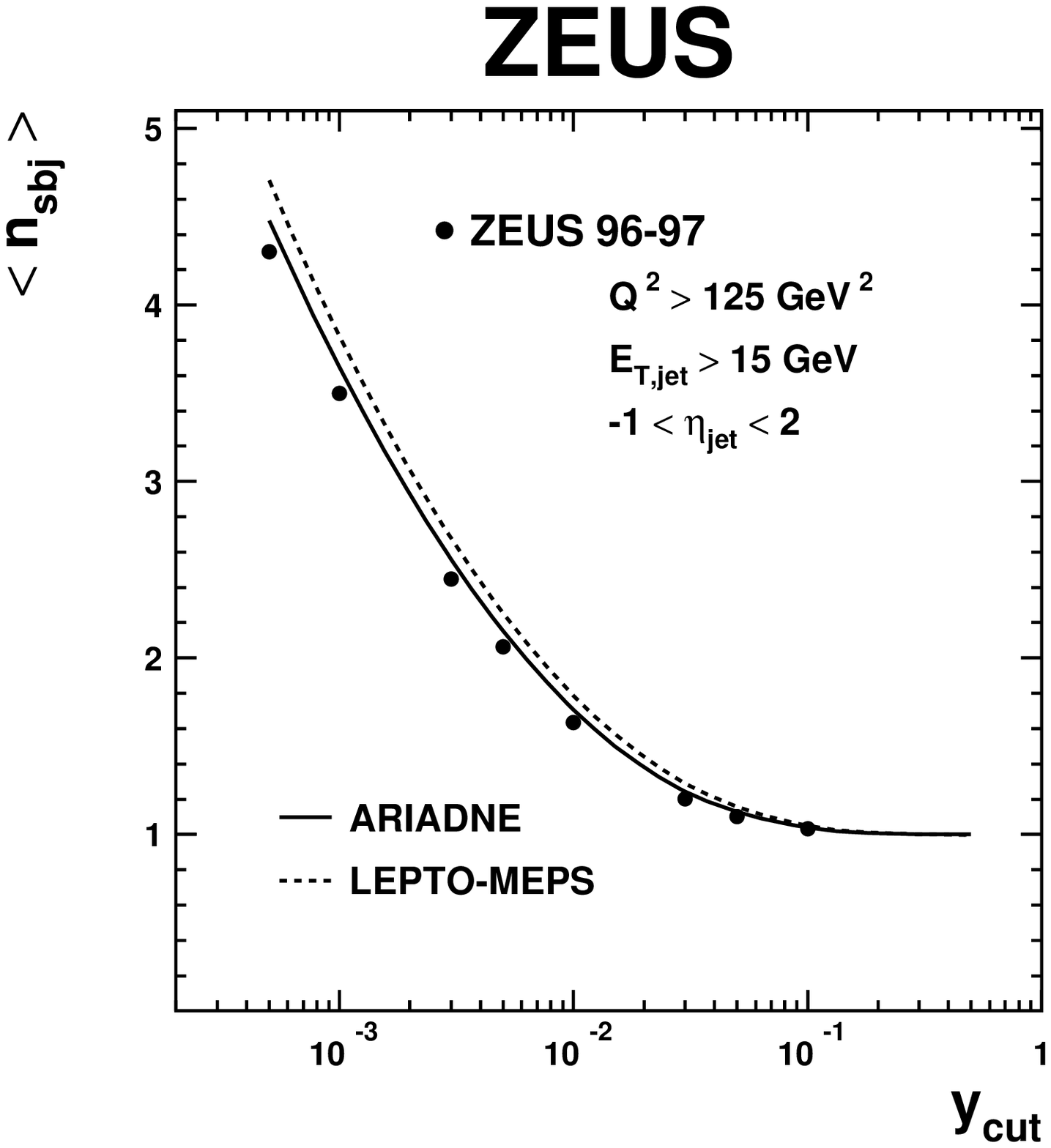,width=14.50cm}}
\put (12.0,18.75){\bf a)}
\put (12.0,7.5){\bf b)}
\put (1.20,0.0){\epsfig{figure=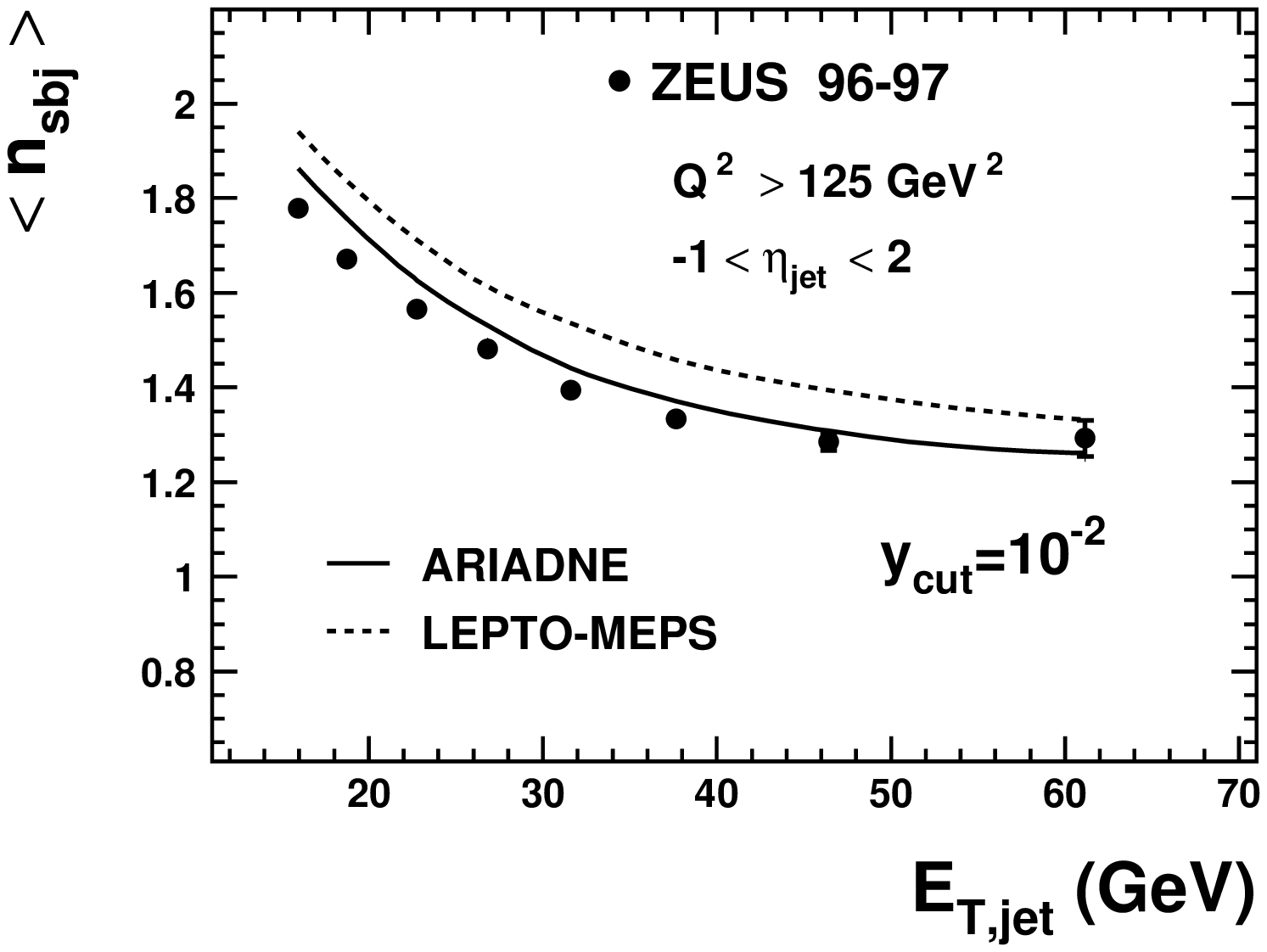,width=14.50cm}}
\end{picture}
\vspace{-1.0cm}
\caption{\label{fig:ycutmc}
The mean subjet multiplicity corrected to the hadron level, $\bigl< n_{\rm sbj} \bigr>$,
as a function of a) $y_{\rm cut}$ and b) $\etjet$ at $y_{\rm cut}=10^{-2}$ for inclusive
jet production in NC DIS with \mbox{$Q^2>125$~GeV$^{\,2}$,} $-1<\etajet<2$ and
$\etjet>15$~GeV~(dots). The inner error bars show the statistical uncertainty.
The outer error bars show the statistical and systematic uncertainties added in quadrature.
For most of the points, the experimental uncertainties are smaller than the size of the
symbols. For comparison, the predictions at the hadron level of the ARIADNE~(solid line)
and
LEPTO-MEPS~(dashed line) models are shown.
}
\end{figure}
\begin{figure}[p]
\setlength{\unitlength}{1.0cm}
\vspace{1.0cm}
\begin{picture} (19.0,16.50)
\put (0.0,0.0){\epsfig{figure=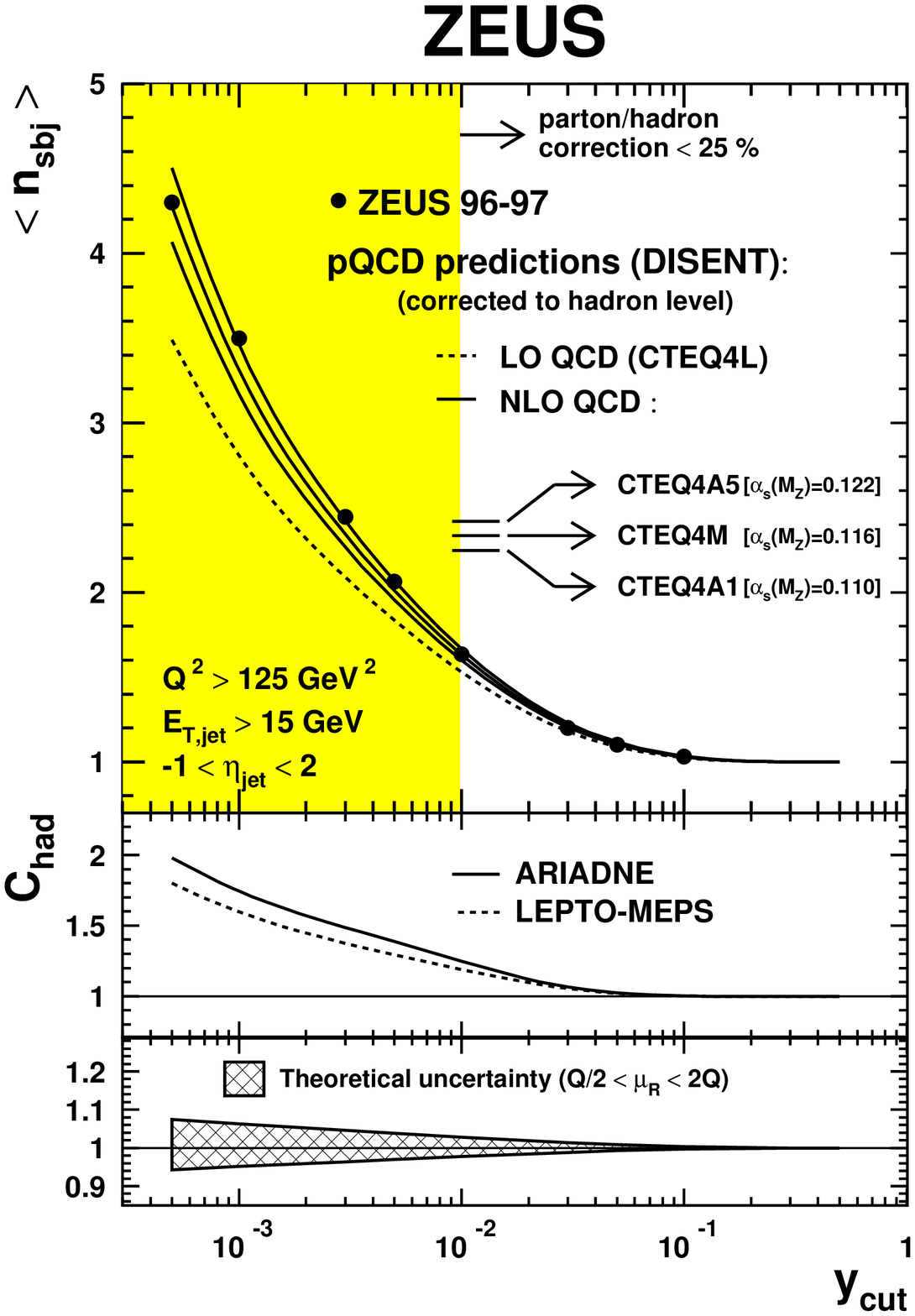,width=16.5cm}}
\put (13.2,17.5){\bf a)}
\put (13.2,7.2){\bf b)}
\put (13.2,4.2){\bf c)}
\end{picture}
\vspace{-1.5cm}
\caption{\label{fig:ycutnlo}
a) The mean subjet multiplicity corrected to the hadron level, $\bigl< n_{\rm sbj} \bigr>$,
as a function of $y_{\rm cut}$ for inclusive jet production in NC DIS with
\mbox{$Q^2>125$~GeV$^{\,2}$,} $-1<\etajet<2$ and $\etjet>15$~GeV~(dots).
The experimental uncertainties are smaller than the size of the symbols. The NLO QCD calculations,
corrected for hadronisation effects and using $\mu_R=\mu_F=Q$, are shown for the CTEQ4 sets of
proton PDFs (CTEQ4A1, lower solid line; CTEQ4M, central solid line; CTEQ4A5, upper solid line). 
The LO QCD calculations, corrected for hadronisation effects and using $\mu_R=\mu_F=Q$ and 
the CTEQ4L set of proton PDFs, are also shown~(dashed line).
b) The parton-to-hadron correction, $C_{\rm had}$, used to correct the QCD predictions and
determined using ARIADNE (solid line) and LEPTO-MEPS (dashed line). c)
The relative uncertainty on the NLO QCD calculation due to the variation of the renormalisation scale.
}
\end{figure}
\begin{figure}[p]
\setlength{\unitlength}{1.0cm}
\begin{picture} (19.0,18.50)
\put (0.0,0.0){\epsfig{figure=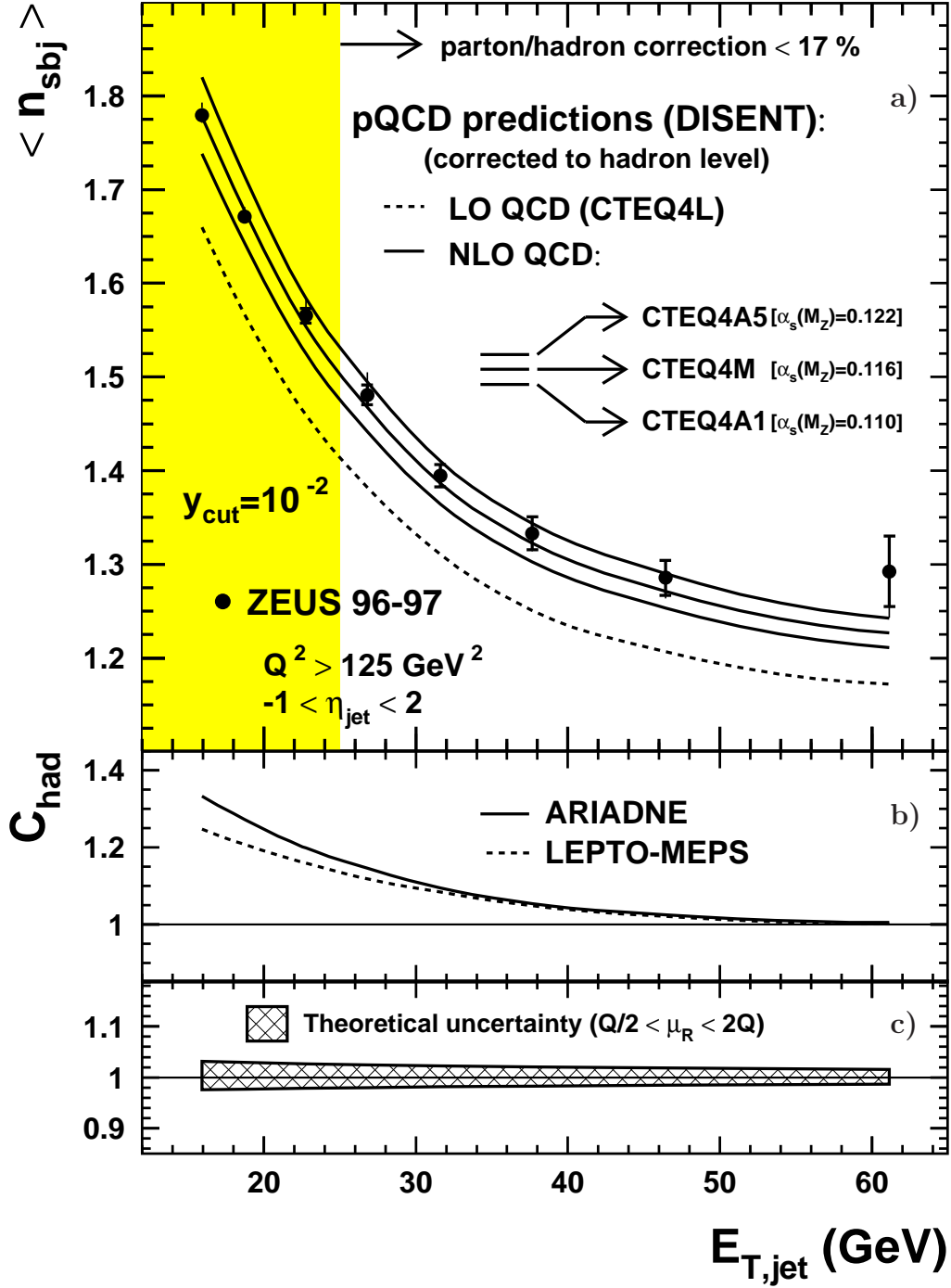,width=16.5cm}}
\put (13.2,17.5){\bf a)}
\put (13.2,7.2){\bf b)}
\put (13.2,4.2){\bf c)}
\end{picture}
\vspace{-1.0cm}
\caption{\label{fig:etnlo}
a) The mean subjet multiplicity corrected to the hadron level, $\bigl< n_{\rm sbj} \bigr>$,
at $y_{\rm cut}=10^{-2}$ as a function of $\etjet$ for inclusive jet production in NC DIS with
$Q^2>125$~GeV$^{\,2}$ and $-1<\etajet<2$~(dots). The inner error bars show the
statistical uncertainty. The outer error bars show the statistical and systematic uncertainties
added in quadrature.
b) The parton-to-hadron correction, $C_{\rm had}$, used to correct the QCD predictions and
determined using ARIADNE (solid line) and LEPTO-MEPS (dashed line).
c) The relative uncertainty on the NLO QCD calculation due to the variation of the renormalisation
scale.
Other details are as described in the
caption to Fig.~\ref{fig:ycutnlo}.
}
\end{figure}